\documentclass[
 aps,
 amsmath,amssymb,
 reprint,%
]{revtex4-2}
\usepackage{graphicx}
\usepackage{dcolumn}
\usepackage{bm}
\usepackage{color,soul}
\usepackage{amsfonts}
\usepackage{mathtools}
\usepackage[utf8]{inputenc}
\usepackage[T1]{fontenc}
\usepackage{mathptmx}
\usepackage{etoolbox}
\usepackage{sidecap}
\usepackage{wrapfig}
\usepackage[dvipsnames]{xcolor}

\newcommand{\g}[1]{{#1}}

\makeatletter
\def\@email#1#2{%
 \endgroup
 \patchcmd{\titleblock@produce}
  {\frontmatter@RRAPformat}
  {\frontmatter@RRAPformat{\produce@RRAP{*#1\href{mailto:#2}{#2}}}\frontmatter@RRAPformat}
  {}{}
}%
\makeatother
\begin{document}

\preprint{AIP/123-QED}


\title[State- versus Reaction-Based Information Processing in Biochemical Networks]{State- versus Reaction-Based Information Processing in Biochemical Networks}
\author{Anne-Lena Moor$^{1,2,*}$}
\author{Age Tjalma$^{3}$}
\author{Manuel Reinhardt$^{3}$}
\author{Pieter Rein ten Wolde$^{3,*}$}
\author{Christoph Zechner$^{4, 1,2,*}$}%
 \email{czechner@sissa.it}
 \email{tenwolde@amolf.nl}
 \email{annemoor96@gmail.com}
\affiliation{ 
$^1$Max Planck Institute of Molecular Cell Biology and Genetics, 01307 Dresden, Germany
}%
\affiliation{ 
$^2$Center for Systems Biology Dresden, 01307 Dresden, Germany
}%
\affiliation{ 
$^3$AMOLF,  Science Park 104, 1098 XG Amsterdam,  Netherlands
}%
\affiliation{$^4$Scuola Internazionale Superiore di Studi Avanzati, 34136 Trieste, Italy}

\date{\today}

\begin{abstract}
Trajectory mutual information is frequently used to quantify information transfer in biochemical systems. Tractable solutions of the trajectory mutual information can be obtained 
via the widely used Linear-Noise Approximation (LNA) using Gaussian channel theory.
This approach is expected to be accurate for sufficiently large systems. However, recent observations show that there are cases, where the mutual information obtained this way differs qualitatively from results derived using an exact Markov jump process formalism, and that the differences remain even in the large copy number regime. 
In this letter, we show that these differences can be explained by introducing the notion of reaction- versus state-based descriptions of trajectories. In chemical systems, the information is encoded in the sequence of reaction events, and the reaction-based trajectories of Markov jump processes capture this information. We show that within the Gaussian formalism, trajectories can be defined either based on individual reaction channels, or on a state-based level, where different reaction channels are summarised into a single noise term. While both definitions agree in terms of copy number fluctuations, state-based trajectories contain in general less information than reaction-based trajectories. The commonly used Gaussian mutual information via the Linear-Noise Approximation is consistent with a state-based trajectory notion, which causes a systematic loss of information independent of system size. We show that an alternative, reaction-based variant of the Gaussian mutual information prevents this loss of information.
We illustrate the consequences of different trajectory descriptions for two common cellular reaction motifs and discuss their connection with Berg-Purcell and Maximum-Likelihood sensing.
\end{abstract}

\maketitle
\paragraph*{Introduction.} 
Accurately responding to time-varying signals is essential for many natural and engineered systems,  ranging from neural and biochemical networks to electronic circuits. The central measure for quantifying information transmission via time-varying signals is the information transmission rate, also called trajectory- or path mutual information rate \cite{shannon1948mathematical, PhysRevLett.102.218101, fano1961transmission, duso2019path, Sinzger2020}.  Mathematically, it is defined as the rate at which the \g{trajectory} mutual information between the input and output signal trajectories increases with the duration of these trajectories in the long-time limit. 
In contrast to other information theoretical measures, such as the instantaneous or time-lagged mutual information \cite{PhysRevE.81.061917}, the \g{trajectory} mutual information not only captures the accuracy of the instantaneous input-output mapping, but also the effect of temporal correlations within the input and output signals.
Cellular signalling networks show that the encoding of information in trajectories can be highly non-trivial. These systems transmit information via chemical reactions between discrete molecules. In the limit that these systems are well mixed, they are appropriately described as discrete Markov jump processes. Yet, it is commonly believed that \g{Gaussian continuum approximations} become accurate when the reactions are linear and the copy numbers are large \cite{PhysRevE.82.031914, 10.1371/journal.pone.0001077, 10.1371/journal.pcbi.1000125}. They can indeed accurately predict the instantaneous mutual information, even when the copy numbers are as small as 10 \cite{PhysRevE.82.031914}.  \g{Consequently}, a Gaussian approach like the Linear-Noise Approximation \cite{vankampen} is frequently used to \g{study information transmission in biochemical systems, where the trajectory mutual information can be conveniently obtained from the  temporal correlation functions of molecule counts \cite{PhysRevLett.102.218101}}. However, it has recently been shown that this Gaussian approach not only significantly underestimates the information rate, but can, in some situations, also lead to qualitatively different behaviour, even when the copy numbers are large and the instantaneous mutual information is correctly predicted \cite{Moor,PhysRevX.13.041017}.  This observation raises the question how information is encoded in trajectories and how this information can be quantified.\\

\begin{figure}[t]
\centering
    \includegraphics[width=0.99\columnwidth]{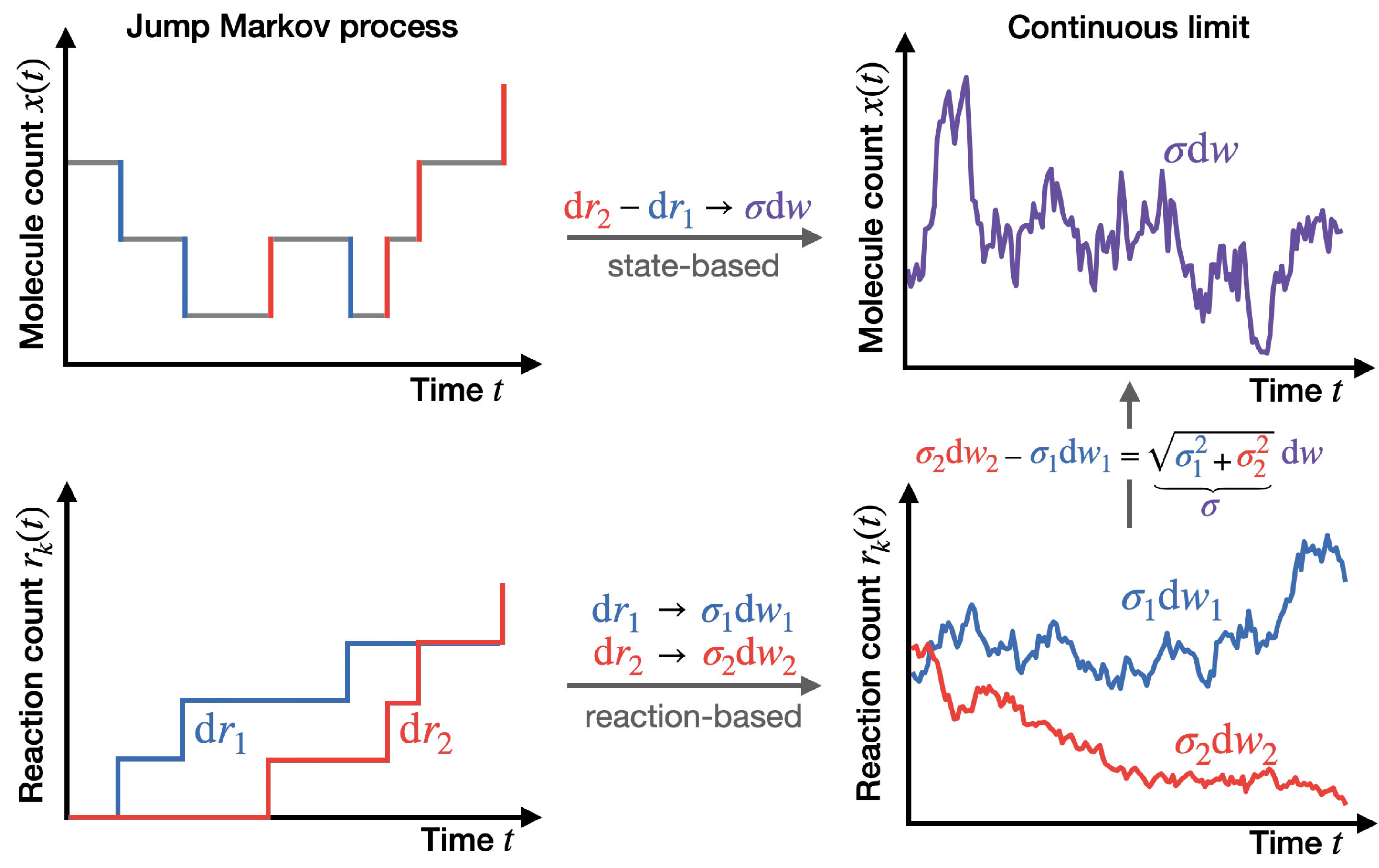}
    \caption{\g{State- versus reaction-based trajectories. As an example, we consider a birth-and-death process with copy number $x(t)$. The discrete trajectory on the left consists of two types of chemical reactions, a degradation reaction with stoichiometric coefficient $s_{x, 1} = -1$ (blue) and a birth reaction with stoichiometric coefficient $s_{x, 2} = 1$ (red). The conventional state-based continuous limit summarises both reactions into effective concentration changes (top), which leads in general to a loss of information in state-based trajectories. In contrast, the suggested reaction-based continuous limit distinguishes birth- and degradation reactions (bottom), thus preserving the information that is encoded in the discrete reaction trajectories.}}
    \label{fig:reacvsconc}
\end{figure}

\g{In this work, we introduce the notion of reaction- versus state-based descriptions of trajectories in the context of information transfer. In chemical systems the information on the input is encoded in the sequence of reaction events of the output, and the discrete trajectories of Markov jump processes capture this encoding. In contrast, the commonly used Gaussian approximation of trajectory mutual information \cite{PhysRevLett.102.218101} entails a state-based trajectory notion, where individual reaction channels are summarised into effective copy number/concentration fluctuations. This step results in a loss of information as we demonstrate using the data-processing inequality \cite{cover1999elements}. This explains why the commonly used form of the Gaussian approximation significantly underestimates trajectory mutual information even in the large copy number regime. This observation naturally leads to an alternative, reaction-based variant of the Gaussian approximation, which explicitly keeps track of the individual reaction channels (see Fig.~\ref{fig:reacvsconc}). We show that this approach accurately predicts the trajectory mutual information for linear chemical systems at large copy numbers. 
Finally, we demonstrate that the distinction between reaction- versus state-based trajectories has ramifications for information transmission in cellular systems, by analyzing systems that can harness the information that resides in the reaction events.}

\paragraph*{Stochastic Reaction Dynamics.} 
We consider a well-mixed reaction system of size $\Omega$, consisting of $M$ chemical species $\mathrm{Z}_1,...,\mathrm{Z}_M$. These species interact via $K$ distinct reaction channels. The state of the system is denoted by the vector $z(t)$ which contains the copy numbers of the respective species at time $t$. To each reaction channel $k$, we associate a rate function $h_k(z(t))$ and define $h(z) = (h_1(z), \ldots, h_K(z))^T$ and $H(z) = \mathrm{diag}(h(z))$. For sufficiently large systems, we can make use of the \g{Linear-Noise} Approximation
$z(t) = \hat{z}(t) + \delta z(t)$, where $\hat{z}(t)$ is the macroscopic mean copy number of $z(t)$ and $\delta z(t)$ is a random fluctuation that scales with the square root of the system size $\Omega^{1/2}$ \cite{vankampen}. For stationary systems, the dynamics of $\delta z(t)$ satisfies
\begin{equation}
\label{eq:lna}
\begin{split}
\mathrm{d}(\delta z(t) ) = S J \delta z(t) \mathrm{d}t  + S  H^{1/2}\mathrm{d}w (t),
\end{split}
\end{equation} 
where $S$ is the stoichiometry matrix with dimension $M\times K$,  $J$ is the Jacobian of $h(z)$ evaluated at the stationary macroscopic mean $z^*$ and 
$\mathrm{d}w$ denotes a $K$-dimensional noise vector of independent, standard Wiener processes, with each element $\mathrm{d}w_k(t)$ reflecting an individual reaction channel  (see \cite{suppl} Section A).

\paragraph*{Reaction- Versus State-Based \g{Trajectories}.}

%
The main difference between a reaction- and state-based \g{trajectories} lies in whether \g{they} are defined using a $K$-dimensional reaction noise vector, or an $M$-dimensional vector describing the effective copy number noise.
In Eq. (\ref{eq:lna}), the matrix $SH^{1/2}$ transforms the $K$-dimensional reaction noise vector $\mathrm dw$ into the $M$-dimensional species-level noise $\mathrm{d}\xi = S H^{1/2} \mathrm{d}w$, summarising the net contribution of reactions to the effective change in copy numbers. Since the linear combination of two \g{Gaussian increments} is again a \g{Gaussian increment}, $\mathrm{d}\xi$ remains a vector of Gaussian increments with a modified covariance structure. For instance, the sum of two independent Gaussian increments $a\mathrm{d}w_1(t) + b\mathrm{d}w_2(t)$ results in a single Gaussian increment $\sqrt{a^2 + b^2}\mathrm{d}w(t)$. This projection leads to the commonly used state-based formulation of the LNA, for which trajectories are defined as $\tilde{z}_0^t = \left\{\mathrm{d}\xi(s) \mid 0\leq s < t\right\}$. 

\g{Importantly, the projection from reaction-based onto state-based increments is irreversible. Summarising reaction channels, as done for state-based \g{trajectories}, can only lead to a loss of information as we demonstrate more explicitly in Case Study I. In fact,}
while the effective noise on a species contains the effect of \textit{all} reactions that modify a species, the information on the input trajectory is, in general, contained only in a subset of reaction channels. The remaining reactions act as a source of uninformative noise which hampers information transfer. 

This observation leads to \g{the formulation of trajectories} at the level of reaction channels. In this formalism, each \g{trajectory} is described by the individual reaction events, i.e. $z_0^t = \left\{\mathrm{d}w_k(s), k\in \{1, ..., K\} \mid 0\leq s < t\right\}$. \g{The reaction-based trajectory} distinguishes individual reaction events and is, hence, analogous to jump processes, where \g{trajectories} are unambiguously defined by the sequence of reaction times and types. In the following, we will show that the previously observed discrepancies between the Gaussian and discrete \g{trajectory} mutual information can be explained by the difference in defining \g{trajectories} in the linear noise regime, i.e. either via the state- or the reaction-based formalism.

\paragraph*{\g{Trajectory} Mutual Information for Gaussian Processes.}
The calculation of mutual information between Gaussian trajectories has been studied extensively \cite{mizakai, HITSUDA197466,f83988fc-492c-3fbc-8bc5-4ac12167fa43}. 
Let $x(t)$ and $y(t)$ be the copy numbers of two chemical species that are part of the system state $z(t)$. The \g{trajectory} mutual information between trajectories of $x(t)$ and $y(t)$ is generally defined as 
\begin{equation}
\label{eq:MI}
I_t^{xy} = \Bigl\langle \log \frac{\mathrm{dP}^{xy}}{\mathrm{d}(\mathrm{P}^x \times \mathrm{P}^y)} \Bigr\rangle , 
\end{equation}
where the term inside the logarithm is the Radon-Nikodym density of the joint path measure $\mathrm{P}^{xy}$ with respect to the product of the marginal path measures $\mathrm{P}^{x} \times \mathrm{P}^y$ \cite{lipster}. 
Using the expressions for the Radon-Nikodym density for Wiener processes, we can evaluate Eq. (\ref{eq:MI}) for both state- and reaction-based \g{trajectories} \cite{lipster, mizakai, HITSUDA197466,f83988fc-492c-3fbc-8bc5-4ac12167fa43}. \g{T}he mutual information between state-based \g{trajectories} $\tilde{x}_0^t \subset \tilde{z}_0^t$ and $\tilde{y}_0^t\subset \tilde{z}_0^t$ reads
\begin{equation}
\label{eq:gaussMIviaRND_cc}
\begin{split}
&\tilde{I}^{xy}_t  = \\ &\frac{1}{2} \int_0^t \frac{1}{\sigma_x^2}  \Bigl(  \langle\mathrm{Var}[ \phi_x(z(s))\mid \tilde{x}_0^s]\rangle  - \langle\mathrm{Var}[ \phi_x(z(s))\mid \tilde{x}_0^s , \tilde{y}_0^s]\rangle \Bigr) \\ &+ \frac{1}{\sigma_y^2}  \Bigl(\langle \mathrm{Var}[\phi_y(z(s))\mid \tilde{y}_0^s]\rangle -  \langle\mathrm{Var}[ \phi_y(z(s))\mid \tilde{x}_0^s , \tilde{y}_0^s]\rangle \Bigr) \mathrm{d}s ,
\end{split}
\end{equation}
where $\phi_x(z(t)) = \sum_{k\in R_{\bar{x}}} s_{x, k}\lambda_k(z(t))$ and $\phi_y(z(t)) = \sum_{k\in R_{\bar{y}}} s_{y, k}\lambda_k(z(t))$ are the drift of $x(t)$ and $y(t)$ with $\lambda_k(z(t))= \sum_{i}\partial_{z_i}h_k(z^*) \delta z_i (t)$, $\sigma_x =\sqrt{ \sum_{k\in R_{\bar{x}}} s_{x,k}^2 h_k (z^*) } $ and $\sigma_y =\sqrt{ \sum_{k\in R_{\bar{y}}} s_{y,k}^2  h_k (z^*) } $ denote the noise magnitude of $x(t)$ and $y(t)$ and $s_{x,k} $ and $s_{y,k} $ are the stoichiometric coefficients of reaction $k$ modifying $x(t)$ and $y(t)$, respectively. The noise magnitudes $\sigma_x$ and $\sigma_y$ are determined by the sets of reactions $R_{\bar{x}}$ and $R_{\bar{y}}$ that  affect $x(t)$ or $y(t)$, respectively. 

Evaluating Eq. (\ref{eq:MI}) for reaction-based \g{trajectories} $x_0^t \subset z_0^t$ and $y_0^t \subset z_0^t$ instead, yields a different equation for the \g{trajectory} mutual information: 
\begin{equation}
\label{eq:gaussMIviaRND_rc}
\begin{split}
I^{xy}_t  &= \frac{1}{2}  \int_0^t  \sum_{k\in R_{x}} \frac{\langle\mathrm{Var}[\lambda_k(z(s))\mid x_0^s]\rangle - \langle\mathrm{Var}[\lambda_k(z(s))\mid x_0^s , y_0^s]\rangle}{\sigma_k^2}  \\ & + \sum_{k\in R_{y}} \frac{\langle \mathrm{Var}[\lambda_k(z(s))\mid y_0^s]\rangle -  \langle\mathrm{Var}[\lambda_k(z(s))\mid x_0^s , y_0^s]\rangle}{\sigma_k^2} \mathrm{d}s ,
\end{split}
\end{equation}
where $\sigma_k =\sqrt{ h_k (z^*) } $ denotes the standard deviation of reaction $k$. The sums in Eq. (\ref{eq:gaussMIviaRND_rc}) run over the subsets $R_{x}\subseteq R_{\bar{x}}$ and $R_{y}\subseteq R_{\bar{y}}$, which contain the reactions that affect $x(t)$ or $y(t)$, respectively, but are caused by species other than $x(t)$ or $y(t)$, as discussed further in Appendix A.  
Note that the rate of information transmission in both cases is given by the integrands of Eq. (\ref{eq:gaussMIviaRND_cc}) and Eq. (\ref{eq:gaussMIviaRND_rc}). 

Comparing Eq. (\ref{eq:gaussMIviaRND_cc}) and Eq. (\ref{eq:gaussMIviaRND_rc}) highlights important differences in the information transfer between state- and reaction-based \g{trajectories}. For state-based \g{trajectories}, the mutual information between species X and Y involves \textit{all} reactions that change the copy numbers of these species. This is reflected in the denominators of Eq. (\ref{eq:gaussMIviaRND_cc}), which contain effective standard deviations $\sigma_x$ and $\sigma_y$ with contributions from all reactions in $R_{\bar{x}}$ and $R_{\bar{y}}$. In contrast, each term in the sums of Eq. (\ref{eq:gaussMIviaRND_rc}) is accompanied by a reaction-specific standard deviation $\sigma_k$.
This is because in the reaction-based description, only subsets of reactions $R_{x}\subseteq R_{\bar{x}}$ and $R_{y}\subseteq R_{\bar{y}}$ are considered, namely those that explicitly contribute to information transfer. For example, in the reaction network $Y \rightarrow X$, $X \rightarrow \emptyset$, both reactions belong to $R_{\bar{x}}$ and contribute to $\sigma_x$, but only the first reaction is also contained in $R_x$, since only that reaction contributes directly to information transfer from Y to X.  Reactions, that are contained in $R_{\bar{x}}$ but not in $R_x$ can contribute to information transfer, but only implicitly via their effect on the expected conditional variances. 
\paragraph*{Calculation of the \g{Trajectory} Mutual Information.} The calculation of the \g{trajectory} mutual information for state- and reaction based \g{trajectories} requires knowledge of the expected conditional variances that appear in Eqs.~(\ref{eq:gaussMIviaRND_cc}) and (\ref{eq:gaussMIviaRND_rc}). The conditional variances inside the expectations are the second moments of a so-called filtering distribution, which captures the statistics of a subset of the species at time $t$, given a complete trajectory of the remaining species between time zero and $t$. For the calculation of $\mathrm{Var}[\lambda_k(z(t))\mid x_0^t]$ in Eq.~(\ref{eq:gaussMIviaRND_rc}), for instance, we need the probability distribution $\pi^x(\bar{z},t)=P(\bar{z}(t)=\bar{z}\mid x_0^t)$, where $\bar{z}(t)$ contains the copy numbers of all species except $x(t)$. \g{Since the considered dynamics are linear and Gaussian, the required conditional distributions and their second moments are tractable. In the case of state-based trajectories, the expected conditional variances are readily available through the well-known Kalman-Bucy filtering equations \cite{Klmn1961NewRI, crisan}. 
In the case of reaction-based trajectories, the calculation of the conditional variances is less straightforward because the standard Kalman-Bucy filter does not apply. We show in \cite{suppl} Sections C-F how tractable solutions of the expected conditional variances can be obtained using filtering theory \cite{KUTSCHIREITER2020102307, crisan}.}

\paragraph*{Case study (I) -- Comparison of Different Network Motifs.}
\g{To illustrate the consequences of using different trajectory descriptions,} we calculated mutual information rates for two common reaction motifs \cite{PhysRevLett.102.218101,TanaseNicola:2006bh} using reaction- and state-based \g{trajectories} (Table \ref{tab:motifs}). Even though the studied motifs are quite simple, the results are markedly different. For motif a, we find that the effect of reaction 3 (i.e. with rate constant $c_3$) is stronger for the reaction-based formalism, where the reaction contributes twice as much to the information rate as in the state-based one. 
Tracking the occurrence of this factor in the calculation of the \g{trajectory} mutual information rate shows that the inclusion of reaction 4 (i.e. with rate constant $c_4$) in the state-based formalism is responsible for the attenuation of reaction 3. 
While only the fluctuations caused by reaction 3 carry information about the signal (species A), in the state-based description, distinguishing reaction 3 from 4 is impossible \g{ as $\mathrm{d}\xi_b(t) = \sqrt{c_3a^* + c_4b^*}\mathrm{d}w_b(t)$}, resulting in a loss of information about the reaction events that encode the signal. Instead, in the reaction-based scheme the fluctuations caused by reaction 3 are tracked explicitly, resulting in a higher mutual information rate. The loss of information associated with state-based trajectories can be explained also by the data-processing inequality \cite{cover1999elements}. 
\g{Since reaction 4 acts purely as a (noisy) post-processing step, with no additional information about the signal $a(t)$, the mutual information rate can only decrease when reactions 3 and 4 are combined (see \cite{suppl} Section H.1 for further discussion)}. 
\\ 
The discrepancies between \g{reaction- and state-based trajectories} can be much more pronounced, as can be seen from motif b. For state-based \g{trajectories}, the noise terms are associated with the net change in copy number of the respective species. In motif b, the reaction-specific noise terms $\sqrt{c_1}\mathrm{d}w_1(t)$, $\sqrt{c_2 a^*}\mathrm{d}w_2(t)$, $\sqrt{c_3 a^*}\mathrm{d}w_3(t)$ and $\sqrt{c_4 b^*}\mathrm{d}w_4(t)$ are projected onto the noise terms of species A and B as $\mathrm{d}\xi_a(t) = \sqrt{c_1 + c_2 a^* + c_3 a^*}\mathrm{d}w_a(t)$ and $\mathrm{d}\xi_b(t) = \sqrt{c_3 a^* + c_4 b^*}\mathrm{d}w_b(t)$, respectively. 
Since reaction 3 affects both $\mathrm{d}\xi_a(t)$ and $\mathrm{d}\xi_b(t)$, these increments are anti-correlated at every moment in time. The \g{trajectory} mutual information for instantly correlated copy number dynamics diverges in the state-based formalism. Using the method by Tostevin \textit{et al.}, we obtain an infinite information rate for motif b \cite{PhysRevLett.102.218101} (see also Ref. \cite{Chétrite_2019}).
In contrast, the reaction-based formalism keeps track of the individual reactions and the information rate remains finite. 
\begin{table*}[htbp]
\centering
\begin{tabular}{l|c|c|c}
Motif & Reaction & State-based rate \cite{PhysRevLett.102.218101} & Reaction-based rate  \\ \hline \hline 
(a) & $A\xrightarrow{c_3} A+B$, $B\xrightarrow{c_4} \emptyset$ & $-\frac{c_2}{2}+ \frac{1}{2}\sqrt{c_2(c_2+c_3)}$ &  $-\frac{c_2}{2}+ \frac{1}{2}\sqrt{c_2(c_2+2c_3)}$ \\ \hline
(b) & $A \xrightarrow{c_3} B \xrightarrow{c_4} \emptyset $ & $ \infty$ & $\frac{c_3}{2} $ 
\end{tabular}
\caption{Stationary mutual information rate between the \g{trajectories} of input species A and output species B of two common reaction motifs obtained via the reaction-based and state-based scheme.  All motifs include the reaction $\emptyset  \xrightleftharpoons[c_2]{c_1} A$. The motifs have been studied earlier in \cite{PhysRevLett.102.218101,TanaseNicola:2006bh}, where motif a is referred to as motif III and motif b as motif II. The state-based information rate has been calculated using the formalism by Tostevin \textit{et al.} \cite{PhysRevLett.102.218101}.}
\label{tab:motifs}
\end{table*} 
\paragraph*{Case study (II) -- Ligand-Receptor Binding.}
Since the reaction-based information rate is, in general, larger than the state-based rate (unless the latter diverges), the question arises whether biological systems are able to access the information stored in individual reactions. To study whether this is possible, we revisit the problem of chemical sensing. In the  mechanism of time integration, first analysed  by Berg and Purcell (BP), the ligand concentration is estimated from the average receptor occupancy over some integration time $T$ \cite{bergpurcell}. This receptor occupancy depends on both the ligand binding and unbinding rate. Endres and Wingreen realised, however, that only the binding events provide information on the concentration, since only the binding rate depends on the concentration \cite{PhysRevLett.103.158101}. This observation led to the scheme of Maximum-Likelihood (ML) sensing \cite{PhysRevLett.103.158101}, in which the concentration is inferred from the mean duration of the unbound state of the receptor, rather than from its average occupancy. The sensing error of this ML scheme is indeed lower than that of the mechanism of time integration, by precisely a factor of 2 \cite{bergpurcell, PhysRevLett.103.158101, bergpurcellreviewed, physicallimits, PhysRevLett.113.148103, PhysRevLett.123.198101,tenWolde:2016ih}. The observation that the information on the ligand concentration is encoded only in the binding reaction, suggests that the ML sensing error is related to the information rate computed for reaction-based signal trajectories. In contrast, the BP sensing error is expected to be related to the information rate of state-based \g{trajectories}. 

\begin{figure}[h]
    \centering
    \includegraphics[width=0.99\columnwidth]{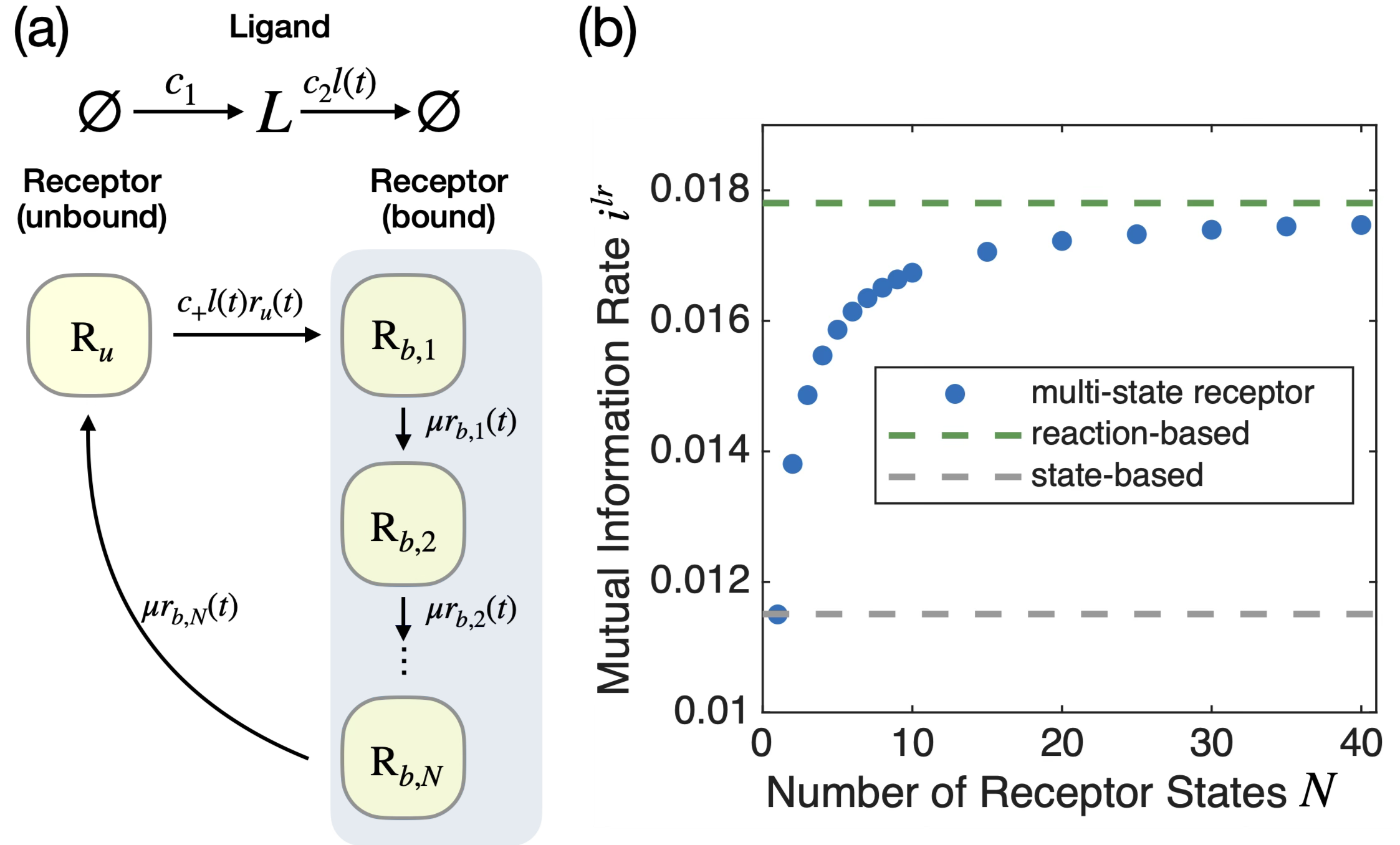}
    \caption{Ligand-receptor binding. (a) Sketch of the multi-state receptor system. The R$_{\mathrm{b}, i}$, $i\in \{1, ..., N \}$ denote the ligand-bound states of the receptor. (b) Stationary mutual information rate $i^{lr}$ between the state-based \g{trajectories} $\tilde{l}_0^t$ and $\tilde{r}_0^t$ of the multi-state receptor dependent on the number of states $N$ (blue dots) in comparison to the mutual information rate of the two state receptor in the state-based description (grey dashed line) and in the reaction-based description (green dashed line). Simulations were performed with parameters $\{c_1, c_2, c_+, c_-, \mu, r_T \} = \{ 1, 0.01, 1, 1, N c_-, 10 \} $. Underlying source code is available at \cite{code}.} 
    \label{fig:ligrec}
\end{figure}

To test this idea, we consider a system consisting of a time-varying ligand concentration and a receptor that senses this ligand by switching between a ligand-bound state and an unbound state (Fig.~\ref{fig:ligrec}a).
We denote by $l(t),r_{\mathrm{b}}(t), r_{\mathrm{u}}(t)$ the copy numbers of the ligand and the ligand-bound and unbound receptors, respectively. \g{ The total number of receptor molecules is given by $r_T = r_{\mathrm{b}}(t) + r_{\mathrm{u}}(t)$ and is constant. In this model, the information rate between the ligand and the ligand-bound receptor state is calculated as}
\begin{equation} 
i^{lr}_{\g{\alpha}} = - \frac{c_2}{2} + \frac{c_2}{2}\sqrt{\frac{\g{\kappa_\alpha} r_T c_- c_+}{c_2 c_- + c_1 c_+} + 1},
\label{eq:iRB}
\end{equation}
\g{with $\alpha=SB$ and $\kappa_{\alpha}=\kappa_{SB}=1$ for state-based trajectories and $\alpha=RB$ and  $\kappa_{\alpha}=\kappa_{RB}=2$ for reaction-based trajectories}.
\g{Interestingly, we can rewrite the expressions for both information rates in terms of the relative sensing error $\eta^2_{\beta} =(\delta l_{\beta} / l^*)^2$, with $\delta l_{\beta}$ referring to the sensing error for Maximum-Likelihood- and Burg-Purcell sensing, i.e. $\beta \in \{ \mathrm{ML}, \mathrm{BP}\}$, respectively and $l^*$ as the stationary average ligand copy number.}
If the downstream system reads out the receptor over a time $T$, then the respective errors per receptor are $\eta^2_{\g{ML}} = 1 / \bar{n}_b$ and $\eta_{\g{BP}}^2 = 2 / \bar{n}_b$,  where $\bar{n}_b = T p / \bar{\tau}_b$ is the average number of binding events during the time $T$, with $p=r_b^*/r_T$ the binding probability, \g{$r_b^*$ the stationary average copy number of bound receptors} and $\bar{\tau}_b=1/c_-$ the average binding time \cite{PhysRevLett.103.158101,bergpurcell,tenWolde:2016ih}. 
 Since the optimal integration time $T$ is on the order of the correlation time $c_2^{-1}$ of the input signal  \cite{10.7554/eLife.62574}, we take $T=c_2^{-1}$. The information rates can then be rewritten as 
\begin{align}\label{eq:iRBML}
    i^{lr}_{\g{\alpha}} &= \frac{c_2}{2}\left( \sqrt{\frac{2 r_T}{\eta_{\g{\beta}}^2l^*}+1} - 1 \right),
\end{align} 
\g{where $\beta = \mathrm{ML} $ for $\alpha = \mathrm{RB}$ and $\beta = \mathrm{BP} $ for $\alpha = \mathrm{SB}$. Hence,}
the respective expressions for the information rate differ only by the sensing error. Since the reaction-based information rate is the maximal information rate, this observation supports the idea that ML sensing can optimally extract the information that the binding reactions provide on the input concentration \cite{PhysRevLett.103.158101}. Noting that for this input the variance $\mathrm{Var}[l(t)]$ is given by the mean $l^*$, we also see  that these expressions lead to the intuitive interpretation that the information rate is determined by the speed at which independent messages are transmitted through the system, set by the correlation time $c_2^{-1}$ of the input, times the number of messages, i.e. concentration levels, that are transmitted reliably per input correlation time, given by $\sqrt{\mathrm{Var}[l(t)]}$ $/ (\delta l_\alpha/\sqrt{r_T})$, with $\delta l_\alpha/\sqrt{r_T}$ the absolute sensing error (\g{for additional details see \cite{suppl} Section I}). 

Still, the question remains whether cells are able to transfer and access as much information as predicted by Eq. (\ref{eq:iRBML}) with \g{$\alpha=RB$}.  Here, we turn to the scheme proposed by Lang \textit{et al.}  \cite{PhysRevLett.113.148103}. In this scheme, the cell estimates the concentration from the average receptor occupancy over some integration time $T$ as in the classical Berg-Purcell scheme. However, while in the classical scheme receptor unbinding is a one-step process, in the scheme of  \cite{PhysRevLett.113.148103} the receptor, upon ligand binding, goes through a sequence of steps before the ligand is released, such that the waiting time for ligand unbinding becomes narrower, thus reducing the contribution from the unbinding noise to the error in the estimate of the receptor occupancy. This leads to the hypothesis that by reading out the receptor \textit{state}, cells are nonetheless able to access the \textit{reaction-based} information rate by eliminating the stochasticity in the unbinding of the ligand.

Calculating the state-based mutual information rate $i^{lr}_{MS}$ between the ligand and the set of all active states of such a multi-state receptor (see \cite{suppl} Section I.4), \g{we find indeed that $i^{lr}_{MS}$ approaches  $i^{lr}_{RB}$ in the reaction-based scheme as $N \rightarrow \infty$} (Fig. \ref{fig:ligrec}b). Increasing the number of receptor states, the unbinding of the ligand becomes \g{less noisy, reducing the information-loss in state-based trajectories.} 
This shows that \g{non-equilibrium reaction cycles} can make the information stored in individual chemical reactions accessible to the cell. 
 In \cite{suppl} Section J, we present a minimal model of a linear multi-state signalling process to support the generality of this statement.
\paragraph*{Conclusion.}
\g{We introduced the notion of state-based and reaction-based trajectories and illustrated their impact on information transfer. 
In the state-based formalism, multiple reaction channels are summarised into effective copy number fluctuations, which, in general, leads to a loss of information about the input signal. The widely used LNA-based approximation of trajectory mutual information relies on state-based trajectories and can therefore substantially underestimate information transmission in chemical systems -- even in the high-copy number regime. We showed that a reformulation of the Gaussian mutual information in terms of reaction-based trajectories prevents this loss of information. This suggests that the principal difference between the state-based Gaussian approximation and the Markov jump process is not the discreteness of molecules, but the encoding of information in reaction events rather than concentrations. 
We demonstrated that the higher information encoded in reaction-based trajectories can be read out through active reaction cycles, which reduce noise stemming from non-informative events. Similar reaction cycles can be found in transduction pathways \cite{PhysRevLett.113.148103}, protein degradation system (e.g., via multistep ubiquitination \cite{pierce2009detection}) and gene promoter networks \cite{suter2011mammalian}. It will be interesting to understand where these systems sit on the spectrum from state- to reaction-based information transfer.}

\g{Our findings are conceptually related to recent work in stochastic thermodynamics, where entropy production is underestimated when a discrete system is taken to the continuum limit under temporal coarse-graining \cite{Busiello_2019a, Busiello_2019b}. Similarly to the present work, these discrepancies are explained by the fact that individual event fluxes are no longer distinguished in the coarse-grained continuous dynamics, leading to a loss of information. Gaining a deeper understanding of how these findings relate will be an interesting topic for future work.}

 \nocite{code}

\begin{acknowledgments}
The authors thank Tommaso Bianucci, Avishek Das and Andreas Hilfinger for useful insights and critical feedback on this work. This work was supported by core funding of the Max Planck Institute of Molecular Cell Biology and Genetics, is part of the Dutch Research Council (NWO) and was partly performed at the research institute AMOLF. This project has received funding from the European Research Council (ERC) under the European Union’s Horizon 2020 research and innovation program (Grant Agreement No. 885065) and was financially supported by NWO through the “Building a Synthetic Cell (BaSyC)” Gravitation Grant (024.003.019).
\end{acknowledgments}
\appendix
\g{\section{Reaction Sets}
In this section, we aim to explain the reaction sets used throughout this letter in detail. We restrict ourselves to the sets $R_z, R_{\bar{z}}, R_{\bar{x}}$ and $R_x$ because $R_{\bar{y}}$ and $R_y$ follow the same rules as $R_{\bar{x}}$ and $R_x$. Let $z(t)$ be a vector containing the state of all species at time $t$. The reaction set $R_z$ contains all reactions that modify any species in $z(t)$. For our filtering approach, we divide the system into an observable part and an unobservable part. Let $x(t) \subseteq z(t)$ be the state of the species we observe. Then, we are interested in the conditional expectation value of all species other than X given the history $x_0^t$ of $x(t)$. In this context, $\bar{z}(t) \subseteq z(t)$ denotes the state of all species other than $x(t)$, i.e. the part of $z(t)$ that we do not observe. The subset $R_{\bar{z}} \subseteq R_z$ is the set of all reactions that modify $\bar{z}(t)$. Analogously, the set $R_{\bar{x}}\subseteq R_z$ is the set of all reactions that modify $x(t)$. For instance, the reactions $X \rightarrow Z_1+X$ and $Z_1 \rightarrow Z_2$ are contained in $R_{\bar{z}}$, whereas the reaction $X\rightarrow \emptyset$ is contained in $R_{\bar{x}}$. As $x(t)$ and $\bar{z}(t)$ can change simultaneously, the intersection of these sets is generally not empty, i.e. $R_{\bar{z}} \cap R_{\bar{x}} \neq \emptyset$. For instance, $X\rightarrow Z_2$ is contained in the intersection of these sets. With $R_x \subseteq R_{\bar{x}}$, we denote the set of reactions that modify $x(t)$ and whose rate depends on a species in $\bar{z}(t)$, such as the reactions $\mathrm{Z}_3 \rightarrow \mathrm{X} + \mathrm{Z}_3$ or $\mathrm{Z}_4 \rightarrow \mathrm{X}$. 
Again, this shows that $x(t)$ and $\bar{z}(t)$ can change simultaneously and it follows $R_{\bar{z}} \cap R_{x} \neq \emptyset$.  Note that we exclude systems involving reactions where (a) the species $X$ and $Y$ are modified simultaneously and (b) the corresponding rate function depends on both $x(t)$ and $y(t)$, such as $\mathrm{Y} + \mathrm{X} \rightarrow \emptyset$.
}

\end{document}